\documentclass[prl,aps,superscriptaddress,twocolumn]{revtex4-1}
\usepackage{siunitx}
\usepackage{graphicx} 
\graphicspath{{graph/}}
\usepackage[usenames]{color}
\usepackage{amsmath,amssymb}
\usepackage{gensymb}
\usepackage{natbib}
\usepackage{xcolor}
\def\strutdepth{\dp\strutbox}
\def\nw#1{\strut\vadjust{\kern-\strutdepth\vtop to0pt{\vss\hbox to\hsize
{\hskip\hsize\hskip5pt$\leftarrow$\hss\strut}}}{\em #1}}

\newcommand{\hb}{\textcolor{black}}
\begin{document}

%\title{Sample article to present \texttt{elsarticle} class\tnoteref{label0}}
%\tnotetext[label0]{This is only an example}
\title{Mechanics and energetics of electromembranes}

\author{Hadrien Bense}
\author{Benoit Roman}
\affiliation{ Physique et M\'{e}canique des Milieux h\'{e}t\'{e}rog\`{e}nes (PMMH), CNRS, ESPCI Paris, PSL Research University, 10 rue Vauquelin, Paris, France ; Sorbonne Universite, Univ. Paris Diderot. }
\author{Jacco Snoeijer}
\affiliation{ Physics of Fluids Group, Faculty of Science and Technology, Mesa+ Institute, University of Twente, 7500 AE Enschede, The Netherlands }
\author{Bruno Andreotti}
\address{ Laboratoire de Physique de l'Ecole Normale Superieure (LPENS), UMR 8023 ENS-CNRS, Univ. Paris-Diderot, 24 rue Lhomond, 75005, Paris.}

\begin{abstract}
The recent \hb{discovery} of electro-active polymers has shown great promises in the field of soft robotics, and was logically followed by experimental, numerical and theoretical developments. \hb{Most of these studies were concerned with systems entirely covered by electrodes}. However, there is a growing interest for partially active polymers, in which the electrode covers only one part of the membrane. Indeed, such actuation can trigger buckling instabilities and so represents a route toward the control of 3D shapes. Here, we study theoretically the behaviour of  such partially active electro-active polymer.
We address two problems:  (i) the electrostatic elastica including geometric non-linearities and partially electro-active strip using a variational approach.  We propose a new interpretation of the equations \hb{of deformation}, by drawing analogies with biological growth, in which the effect of the electric voltage is seen as a change in the reference stress-free state.  (ii) we explain the nature of the  distribution of electrostatic forces on this simple system, which is not trivial. In particular we find that edge effects are playing a major role in this problem.
 %In particular, we use a variational approach to establish the geometrically non-linear electro-mechanical beam equation (similar to Elastica) of a partially covered strip. 
%Lastly, we use a \br{stress-based} approach to interpret the physical origin of the electrostatic forces acting on the system. We notably elucidate the major role played by the edges of the electrode.
%\ba{To be complemented} 
\end{abstract}
\pacs{}

\maketitle
Soft robotics is a new, rapidly developing scientific field. It aims at designing compliant robots able to move in topographically complex environments or able to manipulate fragile objects \cite{Ilievski11,Shepherd11,Martinez13}. Potential applications range from artificial heart~\cite{Roche14} to bio-inspired locomotion \cite{Cheney14,Nawroth2014} or haptic interfaces for virtual reality where the user feels a feedback force from the interface that he manipulates\hb{\cite{Kato07}}.

Electro-Active Polymers offer a promising candidate for building such soft robots. Their large reversible deformations (tens of percent strains) under an applied electric field, triggered a huge interest in academic laboratories over the world \cite{Pelrine2000}. Several types of electro-responsive polymers have been described in the literature, such as anionic, ferroelectrics, liquid-crystalline, electro-rheological~\cite{Brochu10,Kanda11, Bauer14}. In this paper, we are specifically interested in Electro-Active-Polymers (EAP), which are the most simple and inexpensive in terms of manufacturing \cite{Biggs13}. \hb{In their simplest form, EAP} are indeed composed of a thin membrane of elastomer covered by two conductive compliant layers. When an electric field is applied, the electrodes of this soft capacitor are attracted to each other, resulting into in-plane isotropic strain. A wide range of applications of EAPs have been proposed, they include: robots actuation \cite{Kovacs07, Jung07, Carpi07,Carpi05}
%Carpi \& De Rossi 2007)
, rotational motors \cite{Anderson11}, flapping wings \cite{Jordi10,Lau14}, valves \cite{Giousouf13, Murray13}, actuators for biological cells \cite{Akbari12}, Braille displays  \cite{Kato07,Chakraborti12,Vishniakou13}, tunable lenses (\cite{Son12}, \cite{Maffli2015}), artificial chromatophores \cite{Rossiter12,Wang14}, tunable diffracting surfaces \cite{Fang10} or phase shifters for microwave communications \cite{Romano14}. As in the case of piezo-electric ceramics, this electro-mechanical coupling can potentially be reversed in EAPs: an electric power may be harvested from mechanical strain \cite{Kaltseis11, McKay11,Foo12, Zhao14}, or conversely, the device may work as a mechanical sensor \cite{OHalloran08,Sun14}.

In this article we investigate theoretically the behaviour of an electro-active polymer in the case where its \hb{surfaces} are not entirely covered by conductive electrodes. This situation is interesting as it leads to buckling instabilities and is a way to obtain out-of-plane shapes~\cite{Bense17,Li17,Zhu19} through the non-linear response of the structure. The aim of this article is to clarify three questions that arise in this problem. Namely, the literature usually considers membranes completely covered by conductive electrode, and one usually models electrostatic effects by an electrostatic pressure which is twice the actual value of the pressure~\cite{Pelrine2000}, in the case of incompressible elastomers ($\nu=1/2$). In this article we will refer to this description as ``Pelrine's approach''. The argument is based on the  superposition of an  isotropic stress (which has no effect on an incompressible material) to the actual electrostatic pressure on the faces~\cite{Suo10}. The questions we would like to address are: (i) Does this argument apply to cases where electrostatics is only active on a portion of the plate? (ii) \hb{Does the argument remain valid for non-planar buckled states?} (iii) Can we determine the real distribution of electrostatic forces and relate it to the electromechanical equilibrium of a strip? Indeed standard approaches~\cite{Suo10,dorfmann2014} provide a calculation of the mechanical stresses in the material resulting from the electrostatic loading, and even to adequately predict the onset of wrinkling and its possible evolution in pull-in instability \cite{Greaney2018}, but do not provide the real distribution of electrostatic forces from which they originate.  
%A third problem that we wish to address is the exact distribution of electrostatic forces on the material. This issue is crucial to investigate the buckling of an electromembrane. For example a standard approach~\cite{Suo10,dorfmann2014} to study the coupling of electrostatics with mechanics consists in using the Maxwell stress tensor. However, this "stress" is not univocally defined~\cite{dorfmann2014} and is generically non-zero in void regions of space, where there is no charge. 
%$\tau^M$. This particular fields (\ref{MaxwellStress})  obeys $div(\tau^M) + \rho\mathbf{E}=0$, \ba{(CHECK!)} where $\rho$ is the charge density and $\mathbf{E}$ the electric field, and therefore represents the effect of electrostatic forces on charges. Actually, there exists several equivalent choices of this stress field~\cite{dorfmann2014}, as it is defined up to a field with vanishing divergence. Therefore this efficient technique does not provide direct information on the actual electrostatic force distribution. For example the Maxwell stress field may extend in vacuum region in absence of charges (and therefore of forces). 

In this article we use a variational approach to clarify the nature of electrostatic forces acting on an electroactive membrane. Our aim is  to make an explicit link between variational and stress-based approaches of the problem (i.e. a direct description of electrostatic forces on the elastic system), in a detailed way. Particular attention is paid to the edge of the electrode, from the mechanical and the electrostatic point of view. 
We show that the usually \hb{used} Pelrine approach of electro-mechanical coupling extends to these non-uniform activation in the planar case, but that it misses a modification of the bending stiffness due to electrostatic loading. We propose here an alternative analogy, where the electro-active polymer is seen as undergoing a change in its reference stress-free state when submitted to electric voltage. Plasticity, biological growth or chemical swelling are other cases where the reference state of the material evolves, and we suggest that electroactive materials may be seen as analogous to these situations. 

The paper is organized as follows : In section 1, the non-linear out-of plane equilibrium equations for a partially electro-active plate is derived from a variational approach, in the asymptotic limit of thin membranes. These equations involve strain discontinuities at the boundary of the electrodes, and we show how they lead to electrostatically induced buckling. We discuss an analogous description in terms of a modification of the rest-state.
In a second section we turn to stress-based approaches to analyse the previous results in term of force distribution. We discuss the "double electrostatic pressure" description used in a large body of literature on electro-active materials and comment on Maxwell stresses and their interpretation. We finally derive the distribution \hb{of} electrostatic forces directly applied by charges on the elastic plate, and show the importance of electrostatic edge effects, and finally conclude in a last section.
%\js{Need to add structure of the paper (Sec I, II, III).}

\section{Variational approach}

We start by defining the model situation that we will use throughout the article to discuss electroactive coupling. Although we consider here \hb{the} specific plane strain geometry of a strip, and later use specific boundary conditions for the sake of simplicity, the discussion presented is more general.

\subsection{Parametrisation}

\begin{figure}[h]
\begin{center}
\includegraphics{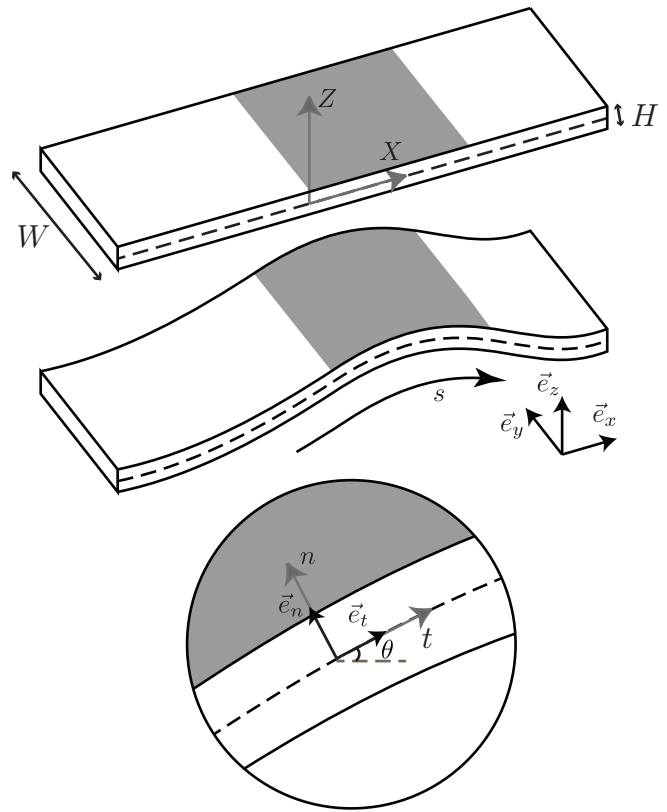} 
\end{center}
\caption{Schematics of the elastic strip partially covered with compliant electrodes. (a) Coordinates of the reference state. (b) coordinates in the current state. \label{fig:sketch}}
\end{figure}

We consider the problem of a thin strip of a dielectric elastic material whose thickness in the reference state  is noted $H$ and whose width is $W$. A portion of the strip has conductive faces \hb{(represented in grey in Fig.\ref{fig:sketch})} which will be submitted to a difference in electric potential $U$ (Fig.\ref{fig:sketch}). \hb{We neglect any stiffening due to the presence of the conductive material}. We start by a general derivation of the equations, and illustrate later (in section variations~\ref{variations}) the buckling for a case with specific boundary conditions (the clamped-clamped case, chosen for simplicity). We restrict ourselves to the situation where the strip is not strained along the transverse $Y$ direction (plane strain configuration). In the reference state the coordinates $X$ and $Z$ respectively run along and across the membrane (Fig.~\ref{fig:sketch}a). We will assume small strain, but follow a geometrically fully non-linear approach, equivalent to Euler's elastica theory~\cite{Audoly10}. This beam geometry, although idealised in comparison with real situations, illustrates the main points of the problem and reveals a non-trivial force distribution.

For the deformed configuration, we adopt a curvilinear description for the centreline position $\mathbf r_0(s)$, where $s$ is the curvilinear coordinate running along the centreline (Fig.~\ref{fig:sketch}b). The local orientation is defined by the tangential and normal unit vectors
\begin{equation}
\mathbf t = \frac{d\mathbf r_0}{ds} = (\cos \theta, \sin \theta), \quad \mathbf n = \frac{d\mathbf t}{d\theta} = (-\sin \theta,\cos \theta).
\end{equation}
The position in the transverse direction is denoted by the coordinate $n$, so that any point inside the membrane is described by 
\begin{equation}\label{eq:transformation}
\mathbf r = \mathbf r_0(s) + n \mathbf n(s) \quad \Rightarrow \quad 
\frac{d\mathbf r}{ds}= (1-n\kappa)\mathbf t , \quad \frac{d\mathbf r}{dn} = \mathbf n,
%d\mathbf r = d\mathbf r_0 + dn \mathbf n + n d\mathbf n = \mathbf t(1-n\kappa) ds + \mathbf n dn,
\end{equation}
where we introduce the curvature $\kappa = d\theta/ds$. From these transformations, one derives the size of an area element as
\begin{equation}\label{eq:areaelement}
d^2\mathbf r = ds dn (1-n\kappa).
\end{equation}
 In the reference state the curvilinear coordinate is noted $S$; it is assumed flat. %so that $\theta(S)=0$. 
 In the slender limit it is useful to introduce the  tangential strain averaged over the cross section of the membrane $e_t$ which determines the stretching
\begin{equation}
ds = (1+e_t) dS,
\end{equation}
where $dS=dX$ is the curvilinear coordinate in the reference state. Similarly, we define the averaged normal strain $e_n(S)$, defined by  
\begin{equation}
h=(1+e_n)H,
\end{equation}
%
%Importantly, this implies that the thickness of the electrically excited membrane is $$.
\hb{Where $h$ is the thickness of the strip in the deformed configuration.\\}
It will turn out convenient to treat the local angle also as a function of the reference coordinate, $\theta(S)$, so that the curvature becomes
\begin{equation}
\kappa \equiv \frac{d\theta}{ds} = \frac{1}{1+e_t} \frac{d\theta}{dS} \simeq  \frac{d\theta}{dS}.
\end{equation}
The approximation in the final step is due to the slender body assumption, i.e. on the small curvature, $(H\kappa)^2 \ll 1$, so that we can neglect terms involving products of both $\kappa$ and $e_t$ or $e_n$   \cite{Audoly10}.

\subsection{Elasticity}
In linear elasticity, the total elastic energy is obtained by integrating over the reference volume as
\begin{equation}\label{eq:elasticenergyfull}
F_e = \frac{E}{(1+\nu)} \int d^3\mathbf X \, \left( \frac{1}{2} \epsilon_{ij} \epsilon_{ij}  + \frac{\nu}{2(1-2\nu)} (\epsilon_{kk})^2 \right),
\end{equation}
where $\epsilon_{ij}$ is the linear strain tensor, defined by:
\begin{equation}
\epsilon_{ij} = \frac{1}{2} \left( \frac{\partial u_i}{\partial X_j} + \frac{\partial u_j}{\partial X_i} \right),
\end{equation}
\hb{and $E$ the Young modulus of the polymer and $\nu$ the Poisson's ratio.\\}
Minimisation of these equations with respect to the displacement field gives the equations of linear elasticity. 

We now follow the standard slender-body reduction of these equations (Euler-Bernouilli assumption), as described for example in \cite{Audoly10}, which gives: 
\begin{equation}\label{eq:membranestrain}
\epsilon_{XX} = e_t - \kappa Z, \quad \epsilon_{ZZ} = e_n + \frac{\nu}{1-\nu} \kappa Z, \quad \epsilon_{XZ}=0, 
\end{equation}
where $e_t$, $e_n$ and $\kappa$ as defined above. We recall that we have assumed a plane strain configuration, so that \hb{$\epsilon_{YY}=\epsilon_{YZ}=\epsilon_{XY}=0$}. In the usual development of the membrane equations, one makes use of the fact that the normal stress $\tau_{ZZ}=0$, which is used e.g. to eliminate the normal strain $e_n$. In the present context, however, this is not allowed since the charged conductors will exert a normal force on the membrane. Hence, we need to retain $e_n$ as a variable. Now, we integrate the elastic energy over the thickness of the layer, inserting (\ref{eq:membranestrain}) in (\ref{eq:elasticenergyfull}), which yields the elastic energy per unit membrane area, 
\begin{eqnarray}
%\int_{-H/2}^{H/2}dY \, W &=&  \int_{-H/2}^{H/2}dY \,  \left[ \left(\frac{1}{2}\lambda + \mu \right) \left( \epsilon_{XX}^2 + e_{YY}^2 \right) + \lambda \epsilon_{XX}\epsilon_{YY} \right] =\nonumber \\
\frac{E}{(1+\nu)} \int_{-H/2}^{H/2} dZ \, \left( \frac{1}{2} \left( \epsilon_{XX}^2 + \epsilon_{ZZ}^2 \right)  + \frac{\nu}{2(1-2\nu)} \left( \epsilon_{XX} + \epsilon_{ZZ} \right)^2 \nonumber
 \right) \\
%&=& \frac{Eh}{2(1-2\nu)(1+\nu)}\left((1-\nu)e^2_t+2\nu e_ne_t + (1-\nu)e^2_n\right) + \frac{Eh^3}{24(1-\nu^2)}\theta'^2 
%\nonumber \\
= \frac{1}{2}{\cal Y}\left(e^2_t+\frac{2\nu}{1-\nu} e_ne_t + e^2_n\right) + \frac{1}{2}B\theta'^2.\nonumber
\end{eqnarray}
where the prime is used for the derivative with respect to $S$. We have introduced the effective stretching modulus ${\cal Y}$ and the bending modulus $B$, defined as:
\begin{equation}
{\cal Y} = \frac{(1-\nu) EH}{(1-2\nu)(1+\nu)} \quad{\rm and } \quad B = \frac{EH^3}{12(1-\nu^2)}.
\end{equation}
Finally, the elastic energy per unit width, which is a functional of the fields $e_t(S)$, $e_n(S)$ and $\theta(S)$, reads:
\begin{equation}\label{eq:elasticreduced}
F_e = \int dS \,  \left[ \frac{1}{2}{\cal Y}\left(e^2_t+\frac{2\nu}{1-\nu} e_ne_t + e^2_n\right) + \frac{1}{2}B\theta'^2 \right].
\end{equation}
The first term represents the in-plane energy due to tangential and normal strain, while the second term represents the plate bending energy.
%
%\ba{CHECKED WITH OLD NOTES:  In the framework of linear elastic stress-strain relationship with Young modulus $E$ and Poisson ratio $\nu$,  the elastic energy reads :
%%
%\begin{equation*} 
%F_e=\int_0^b \frac{dS}{2} \frac{Ehw}{(1+\nu)(1-2\nu)} [ (1-\nu)e_t^2 + 2\nu e_t e_n + (1-\nu)e_n^2 ] + \int_0^b\frac{dS}{2} \frac{Eh^3w}{12(1-\nu^2)} \left( \frac{d\theta}{dS}\right)^2 
%\end{equation*}
%%
%where the first terms represents the in-plane membrane energy, whereas the second term corresponds to bending energy of the plate.}

\subsection{Electrostatics}
The two conductive electrodes form a capacitor maintained at constant potential difference $U$. Denoting by $C$ the capacitance, the electrostatic energy reads $\frac12 C U^2$. Elastic deformations of the membrane lead to a change in the capacitance by $\delta C$. The source that maintains the electrical potential difference $U$ constant brings a charge $U\delta C$ to the capacitor, and as such the capacitor receives a work $U^2 \delta C$ performed by the source. By consequence, at constant potential, the total electrostatic free energy to be 
minimised~\cite{Landau} is $F_U= -\frac12 CU^2$. %\js{add reference, e.g. LandauLifshtitz electro, or FeynmannLectures}
To explicitly perform the minimisation with respect to the elastic degrees of freedom, it is convenient to express the energy in terms of the electrostatic potential $\phi$. Namely, we can then write as a volume integral over the current state
\begin{equation}\label{eq:fuvolumetric}
F_U = - \frac{1}{2} \int d^3\mathbf x \, \varepsilon|\nabla \phi|^2,
\end{equation}
Where $\varepsilon$ is the permitivity of the membrane. We anticipate that $\nabla \phi=0$ everywhere outside the membrane, which implies that we neglect the energetic contribution due to fringe effects near the edge of the plate. Hence, the integral in (\ref{eq:fuvolumetric}) runs over the volume of the membrane, in the current state. Minimisation of the electrostatic energy with respect to the potential $\phi$ yields the expected field equation $\nabla^2 \phi=0$, to be solved inside the deformed membrane.

Within the same slender body asymptotics ($(h \kappa)^2 \ll 1$) as used for the strain field, the gradient of the potential is approximately constant inside the thin gap and points normal to the surface. We therefore seek a solution of the form
\begin{equation}\label{eq:expansionphi}
\phi = \frac{U }{h} \left(n+\frac{h}{2}\right)+ (h \kappa) \phi_1(n,s),
\end{equation}
where $n$ is the local coordinate normal to the surface and we recall that $h=(1+e_n)H$\hb{, and $\phi_1(n,s)$ a function that remains to be determined}. Note the similarity with the expansion (\ref{eq:membranestrain}) for the strain field, which also consists of a solution corresponding to a flat membrane complemented with a curvature correction. 

To obtain explicit expressions, we now need to express the gradient $\nabla \phi$ and the Laplacian $\nabla^2 \phi$ in terms of the curvilinear coordinates $s$ and $n$. This can be achieved formally through the transformation defined by (\ref{eq:transformation}), which  gives
\begin{eqnarray}
\nabla \phi &=& \frac{\partial \phi}{\partial n} \mathbf n + \frac{1}{1- n\kappa} \frac{\partial \phi}{\partial s} \mathbf t, \\
\nabla^2 \phi &\simeq&  \frac{\partial^2 \phi}{\partial n^2} - \kappa \frac{\partial \phi}{\partial n},
\end{eqnarray}
where for the Laplacian we retained only the leading order term in $\kappa$; this term originates from $\mathbf t \cdot \partial \mathbf n/\partial s =-\kappa$. We now insert the expansion (\ref{eq:expansionphi}) in $\nabla^2 \phi=0$, which gives to leading order:
\begin{eqnarray}
\nabla^2 \phi &\simeq& - \kappa \frac{U}{h}  +  (h \kappa) \frac{\partial ^2\phi_1}{\partial n^2} =0.
\end{eqnarray}
This equation governing the curvature correction $\phi_1$ can be integrated twice to yield 
\begin{equation}
\phi = U \left[ \frac{n+\frac{h}{2}}{h} + \frac{1}{2} \frac{(n+\frac{h}{2})(n-\frac{h}{2})}{h^2} \left( h\kappa \right)    \right], \label{eqpotential}
\end{equation}
where we imposed boundary conditions $\phi=U$ and $0$ respectively at $n=h/2$ \hb{and $n=-h/2$ respectively}.
%
%
%\textcolor{red}{JS: alternative route}
%
%
%Now, we intend to perform the same slender reduction as used for the strain field. For this, we assume the thickness $h$ to be locally uniform, while the membrane is weakly curved $h \theta' \ll 1$. We then use the locally curvilinear coordinates $(s,n)$, for which the Laplacian takes the form \textcolor{red}{js: check if this exact, or only to leading order}
%
%\begin{equation}\label{eq:laplcurv}
%\nabla^2 \phi = \frac{\partial^2 \phi}{\partial n^2} +  \theta' \frac{\partial \phi}{\partial n} +  \frac{\partial^2 \phi}{\partial s^2}.
%\end{equation}
%\textcolor{red}{js: maybe in $x,y$ to mimic the elastic derivation in $X,Y$.}
%We then expand the potential as
%
%\begin{equation}
%\phi(s,n)= \phi_0(n) + (h\theta')  \phi_1(n).
%\end{equation}
%Inserted in (\ref{eq:laplcurv}) we can solve up to the leading order in $h \theta'$. Imposing the boundary conditions $\phi(s,0)=0$ and $\phi(s,h)=U$, this gives
%%
%\begin{equation}
%\phi(s,n) = \frac{U}{h} \left[ \frac{n}{h}- \frac{1}{2} \frac{n(n-h)}{h^2} \left( h\theta' \right)    \right].
%\end{equation}
%
Now that we have solved for the field inside the membrane, including the correction due to curvature, we can evaluate the volumetric energy density 
\begin{equation}
-\frac{\varepsilon}{2} |\nabla \phi|^2 = 
- \frac{\varepsilon}{2} \frac{ U^2}{h^2} \left[1 + n\kappa \right]^2.
\end{equation}
%
%\begin{equation}
%\frac{\epsilon}{2} |\nabla \phi|^2 = 
%\frac{\epsilon}{2} \frac{ U^2}{h^2} \left[1 + \frac{2n-h}{h} \left( h\kappa \right) + 
%\frac{1}{4} \frac{(2n-h)^2}{h^2}  \left( h\kappa \right)^2   \right].
%\end{equation}
%
Once again, we integrate this across the membrane, but we need to bear in mind that the area element also contains a curvature correction, $d^2\mathbf r = ds dn (1-n\kappa)$, as derived in (\ref{eq:areaelement}). With this, we derive the electrostatic energy density per unit surface of the membrane:
\begin{equation}
-\int_{-h/2}^{h/2} dn (1-n\kappa)\, \frac{\varepsilon}{2} |\nabla \phi|^2 
= - \frac{\varepsilon}{2} \frac{ U^2}{h}\left[ 1 - \frac{1}{12} \left( h\kappa \right)^2   \right] \label{eqcurvecapa} 
\end{equation}
It is worth noting that the same expression can be recovered from the capacity of the capacitor formed by two coaxial cylinders, which reads:
\begin{equation}
C=\frac{2\pi \varepsilon w}{\ln\left(\frac{1+\kappa h/2}{1-\kappa h/2}\right)}.
\end{equation}
Indeed, the electrostatic energy per unit perimeter~$-\frac{1}{2} C U^2 / (2\pi/\kappa)$ when expanded for small curvature $\kappa h \ll 1$ leads to equation~(\ref{eqcurvecapa}).

%Indeed it suffices to expand for small curvature $\kappa H \ll 1$ the electrostatic energy of this cylindrical capacitor per unit perimeter~$- C U^2\kappa/(4\pi)$.

% 
We thus find the ``slender" expression for the electrostatic energy (per unit width of the plate):
\begin{equation}
F_U = -   \int ds \, \frac{\varepsilon}{2} \frac{ U^2}{h}\left[ 1 - \frac{1}{12} \left( h\kappa \right)^2   \right].
\end{equation}
The first term represents the energy stored in a plane capacitor with conductors separated by a distance $h$. It is proportional to the surface --~we recall that $F_U$ is written per unit membrane width~-- and is therefore analogous to a  surface tension~\cite{deGennes} but with a negative sign\hb{, similarly to what is observed in electrowetting \cite{Lippman}}. 

Moreover, in analogy to the elastic energy (\ref{eq:elasticreduced}), the electrostatic energy represents a bending rigidity as can be seen from the term $\sim \kappa^2$. The associated electrostatic bending modulus is
\begin{equation}
B_U = \frac{\varepsilon U^2h}{12} \simeq  \frac{\varepsilon U^2H}{12}, \label{eqBu}
\end{equation}
where in the remainder we will keep the leading order in strain $(h \simeq H)$. This electrostatic stiffening may also be computed using Maxwell stress tensor (see Appendix\ref{AppendixBU}).

Electrostatics is naturally expressed in Eulerian coordinates. When combined with the elastic formulation, however, it is important to bring this to Lagrangian coordinates. Using the transformations discussed above, this gives
\begin{equation}
\label{electroenergy}
F_U = -  \frac{\varepsilon}{2} \frac{ U^2}{H} \int dS\,  \frac{1+e_t}{1+e_n} + \int dS \, \frac{1}{2} B_U \theta'^2   .
\end{equation}
Like the elastic energy (\ref{eq:elasticreduced}), $F_U$ is now a functional of $e_t(S)$, $e_n(S)$, $\theta'(S)$. Note that for the bending term we neglect higher order corrections in $e_t$, $e_n$ and $\kappa H$.

\subsection{Variations} \label{variations}
For the sake of simplicity we choose here specific boundary conditions. We assume that the position of both ends of the strip are fixed. In particular, the end point position, measured from the origin reads:
\begin{equation}
\mathbf r_{\rm end} = \int dS \, (1+e_t) \mathbf t .
\end{equation}
This constraint will be imposed with a Lagrange multipler, $\mathbf f$, which in fact represents the load applied at the end of the plate. The total energy to be minimised then reads
\begin{eqnarray}
F &=& F_e + F_U = 
\int dS \, \mathcal F(e_t,e_n,\theta) ,
\end{eqnarray}
with the surface density of energy given by
\begin{eqnarray}
\mathcal F(e_t,e_n,\theta)  &=& 
\frac{1}{2}{\cal Y}\left(e^2_t+\frac{2\nu}{1-\nu} e_ne_t + e^2_n\right) + \frac{1}{2}(B+B_U) \theta'^2 \nonumber \\
&& -  \frac{\varepsilon}{2} \frac{ U^2}{H}  \frac{1+e_t}{1+e_n} -
  (1+e_t) \mathbf f \cdot  \mathbf t,
\end{eqnarray}
where this expression must be used with $B_U=0$ and $U=0$ in the portion of the strip that is not covered by the electrode.

The equilibrium equations are obtained by variation of $\theta$, $e_t$, $e_n$. This gives the set of equations
\begin{eqnarray}
\frac{\delta F}{\delta \theta(S)} &=& - (B+B_U) \theta'' - (1+e_t) \mathbf f \cdot \mathbf n=0, \label{eqbend}\\
\frac{\delta F}{\delta e_n(S)} &=& {\cal Y} \left( e_n + \frac{\nu}{1-\nu} e_t \right)  + \frac{\varepsilon}{2}\frac{U^2}{H} \frac{1+e_t}{(1+e_n)^2}=0, \label{eqen} \\
\frac{\delta F}{\delta e_t(S)} &=& {\cal Y} \left( e_t + \frac{\nu}{1-\nu} e_n \right)  - \frac{\varepsilon}{2}\frac{U^2}{H} - \mathbf f \cdot \mathbf t=0 \label{eqet}.
\end{eqnarray}
The terms involving $B_U$ and $U$ are only present in the area covered by electrodes. \hb{These equations are valid for large displacement as they include the complete geometric non-linearities and can therefore be applied to realistic electro-active polymers. In the following, we will for simplicity consider the small strain limit and replace $1+e_n \simeq 1$ and $1+e_t\simeq 1$.}

%Although this is not apparent in equation~(\ref{eqtorque}) }, we will see below that the tension $\mathbf f$ contains electrostatic contributions. 
Equations~(\ref{eqet}-\ref{eqen}) are in fact a force balance projected along the normal and tangential direction, and provide the equations governing the normal  and tangential strain, 
\begin{eqnarray}
{\cal Y} \left( e_n + \frac{\nu}{1-\nu} e_t \right)  &=& - H {\mathcal P}, \label{eqEt}\\
T \equiv \mathbf f \cdot \mathbf t &=& {\cal Y} \left( e_t + \frac{\nu}{1-\nu} e_n \right)  - H {\mathcal P}. \label{eqEn}
\end{eqnarray}
Here we defined $T \equiv \mathbf f \cdot \mathbf t $ as the membrane tension (induced by the imposed force at the edge), and ${\mathcal P}$ as the electrostatic pressure, defined as the usual attractive pressure between two parallel conducting plates 
\begin{equation}
{\mathcal P} =  \frac{\varepsilon}{2}\frac{U^2}{H^2}.
\end{equation}
%
%Note that Eq.(\ref{eqEt}) shows that the normal stress $\sigma_{ZZ}=-{\mathcal P}$ leads to some compression ($e_n<0$)}
%
%\br{THERE WAS A MISSING $^*$ :
%\begin{equation}
%{\cal Y} e_n = - H{\mathcal P} - \frac{\nu}{1-\nu} {\cal Y} e_t.
%\end{equation}
%}
%
%This expression can be inserted in the equation for the tangential strain (\ref{eqet}), $\delta F/\delta e_t(S)=0$, which then gives
%
%T \equiv \mathbf f \cdot \mathbf t =\frac{1-2\nu}{(1-\nu)^2}{\cal Y} e_t  - \frac{1}{1-\nu} H {\mathcal P}
Usually, in plate or beam mechanics, one is not interested in the normal strain. We therefore eliminate $e_n$  and obtain the equation governing the strain $e_t$:
\begin{eqnarray}
T &=& \frac{EH}{1-\nu^2} e_t - \frac{1}{1-\nu} H {\mathcal P}. \label{eqtanForce}
\end{eqnarray}
The total tension consists of the elastic tension of the membrane (first term), and a compressive (negative) tension induced by electrostatics (second term). While this equation for $e_t$ is the most important for the beam, we will for later reference also  compute the normal strain $e_n$, which reads
\begin{equation}
\frac{EH}{1-\nu^2} e_n + \frac{1}{1-\nu} H {\mathcal P} = - \frac{\nu }{1-\nu}T  \label{eqenseul}.
\end{equation}
Again, equations (\ref{eqtanForce},\ref{eqenseul}) are also valid in the regions that are free of electrode, by using ${\mathcal P}=0$.

Interestingly, at the transition between the conductive zone and the non covered membrane, ${\mathcal P}$ exhibits a discontinuity since the electrostatic pressure ${\mathcal P}$ drops to zero outside the electrodes. But since $T=\mathbf f \cdot \mathbf t$ is continuous, this implies that there must be a discontinuity of the strain $e_t$ and $e_n$ at the boundary of the electrode. Following Hooke's law, this also implies a discontinuity of the elastic stress, whose physical origin will be  discussed in sections Force transmitted by surface charges~\ref{mechinterp} and Edge effects~\ref{mechinterp_sing}. %\js{check referencing after subsections are reorganised.} %\br{WE SHOULD say more importantly that the stress is discontinuous also.}

We now turn our attention to the bending equation~(\ref{eqbend}). In the small strain approximation, it then reduces to the standard elastica equation~\cite{Love},
\begin{equation}
(B+B_U) \theta'' +  \mathbf f \cdot \mathbf n =0, \label{eqtorque}
\end{equation}
where the only modification due to electrostatics is the additional bending rigidity $B_U$, present in the part covered by electrodes. We also find that the torque $(B+B_U)\theta'$ is continuous across the edges of the electrode and therefore that the curvature $\theta'$ is either null or discontinuous at this boundary. It is instructive to estimate here the relative magnitude of the electrostatic added stiffness $B_U$ by noting that $B_U/B = 2 {\mathcal P} / E(1-\nu^2)$ is on the order of the strains ($e_n,e_t$) in the material, \hb{assumed to be small in the previous derivation.} However, for practical cases where the stretching might be significant, the electrostatic bending rigidity needs to be taken into account.

As a summary the equilibrium equations for this elecrostatically activated beam are 
\begin{eqnarray*}
(B+B_U) \theta'' +  \mathbf f \cdot \mathbf n &=&0  \\
T \equiv \mathbf f \cdot \mathbf t &=& \frac{EH}{1-\nu^2} e_t - \frac{1}{1-\nu} H {\mathcal P}.
\end{eqnarray*}
where we emphasise that these equations are valid everywhere on the strip, taking $(B_U=0,{\mathcal P}=0)$ on parts that are not covered by electrodes. However, we remind the reader that the discontinuities in ${\mathcal P}$ and $B_U$ at the boundaries of the electrode lead to discontinuities of strains $e_t$, $e_n$ and curvature $\theta'$.

\subsection{Equivalent growth \hb{model}} \label{section_eqgrowth}
An alternative way to write the equilibrium for the strain (\ref{eqtanForce},\ref{eqenseul}) is the form 
\begin{eqnarray}
\label{eqtanForce2}
 \frac{EH}{1-\nu^2} (e_t -e_t^0)  &=& T . \\
\frac{EH}{1-\nu^2} (e_n -e_n^0) &=& - \frac{\nu }{1-\nu}T,
\end{eqnarray}
with
%  (rather than $\frac{1}{1+\nu}\frac{{\mathcal P}}{E})$: }
\begin{equation}
e_t^0= (1+\nu)\frac{{\mathcal P}}{E}; \hspace{1cm} e_n^0= -(1+\nu)\frac{{\mathcal P}}{E}.
\end{equation}
{Written in this way, these equations allow another interpretation of the electrostatic loading: applying a voltage is equivalent to defining a new reference state $(e_t^0,e_n^0)$, where by reference state we mean the state attained when there is no external mechanical loading $T$.}%These equations can be interpreted as if there was a new reference state $(e_t^0,e_n^0)$, where by reference state we mean the state attained when there is no external mechanical loading $T$. 
This situation is similar to cases of inelastic strain, such as those produced by growth or plastic deformation, which effectively redefines the reference state. 
Hence, the electrostatic loading can be seen as a modification of the rest length of the strip, and providing a new bending rigidity $B+B_U$. 

\subsection{Geometrically non-linear equations and buckling instability}
To illustrate how the electrostatic loading enters the equations, let us for example study a simple but geometrically non-linear problem, namely the electrostatically induced buckling instability where the membrane has its two ends clamped. Under an imposed potential $U$ on the electrode, the straight solution $\theta(S)=0$ remains a solution. For this initially flat state $\mathbf t =(1,0)$ and (\ref{eqtanForce}) implies that $e_t$ is piecewise constant. Imposing the constraint of no edge displacement, then implies $\int e_t dS = e_{t,U} L_U + e_{t,0} (L-L_U)=0$, where $L_U$ is the portion of the membrane that is covered by the electrode and $L$ is the total length. Then, (\ref{eqtanForce}) implies that the flat state comes with a compressive tension 
\begin{equation}
%\mathbf f = - \frac{1}{1-\nu} H {\mathcal P} \mathbf t. MODIFIED BY AREA COVERED BY ELECTRODE
T = - \frac{1}{1-\nu} H {\mathcal P} \frac{L_U}{L}.
\end{equation}
We see that the magnitude of this compressive force increases with the potential $U$, and depends on the fraction $L_U/L$ that is covered by the electrode. 
Within the viewpoint presented in~section equivalent growth model~\ref{section_eqgrowth}, this compression has a very simple interpretation, 
as the strip can be seen as taking a new rest state which becomes longer (and stiffer) in the electrode area; for fixed edges this implies a global compression.
The linear stability analysis of (\ref{eqtanForce}) and (\ref{eqtorque}) around this solution shows that this system is unstable under classical buckling (see~\cite{Bense17} for a study of an axisymmetric case for this buckling problem). 
For the simplifying case where the entire membrane with length $L$ is covered with the electrodes,  the electrostatic pressure, and thus $e_t$ and $B_U$, are uniform along the membrane. The buckling threshold can be identified for small perturbations of the elastica equation, which becomes
\begin{equation}
\theta'' +  \frac{H{\mathcal P}}{(1-\nu)(B+B_U)} \theta =0. \hspace{1cm}  \theta(0)=\theta(L)=0
\end{equation}
Defining Euler's critical load $T_c=4\pi^2 (B+B_U) /L^2$ for clamped-clamped boundary conditions, we thus find
\begin{equation}
 \frac{H{\mathcal P}}{(1-\nu)(B+B_U)} =4 \pi^2/L^2.
\end{equation}
This determines the critical electrostatic pressure $\mathcal P$, and thus the critical potential $U$, for buckling.

%\br{CUT}
%Considering the incompressible limit $\nu \to 1/2$, one obtains an expression of the tension $T = \frac43 EH e_t - 2 H {\mathcal P}$ similar to that obtained by Pelrine in the planar case \cite{Pelrine98} and generalised here to the situation where the rod is bent. It is equivalent to the force which would be obtained if a mechanical pressure twice as big as the Maxwell stress was exerted on an infinite rigid plate capacitor. This striking behaviour is recovered in the limit where $\nu$ vanishes, for which the tangential stress decouples from the normal stress. Still, the tension reads $T=EH  e_t - H {\mathcal P}$ and increases as ${\mathcal P}$. In the next section, we will provide an interpretation of this result in terms of mechanics.

\section{Stress-based approach}
We have established above the conditions of equilibrium from variational principles, offering a systematic and rigorous route to describe the state of the membrane. Now we provide a detailed  interpretation of these results in term of stresses (mechanical point of view), and point out various subtleties associated to finding an accurate description in terms of a balance of forces. For the sake of clarity, reasonings will be performed in the simple case of a flat membrane.

\subsection{Pelrine's point of view: Can the electrostatic coupling be reduced to a homogeneous pressure?}

We start by evaluating a common approach found in the literature on incompressible, electro-active elastomers (Poisson ratio $\nu=1/2$)  \cite{Pelrine98}: it is often assumed that the electrostatic effect can be entirely reproduced by considering a pressure $2{\mathcal P}$ exerted by the electrode on the membrane, which is twice as large as the electrostatic pressure acting on a rigid plate capacitor. Considering a flat membrane, the electrostatic free energy per unit width is proportional to the \hb{covered} zone $l$ and inversely proportional to the thickness $h$:
\begin{equation}
F_U = -   \frac{\varepsilon}{2} \frac{ U^2 l}{h}
\label{electroenergyflat}
\end{equation}
The variation of the free energy $F_U$ with respect to the two degrees of freedom $h$ and $l$,
\begin{equation}
dF_U = \frac{\partial F_U}{\partial h} dh+ \frac{\partial F_U}{\partial l} dl
\label{varelectroenergyflat}
\end{equation}
reveals the resultant of the forces exerted on the electrode. The variation with respect to $h$ gives a force normal to the electrode and equal to ${\mathcal P}l$ while the variation with respect to the length $l$ gives a negative tension equal to $ -  {\mathcal P} h$ which must be a force parallel to the electrode. Noting that the free energy (\ref{electroenergyflat}) is proportional to the surface, this force is analogous to a negative surface tension. {Note that volume conservation implies that the two degrees of freedom are linked by $h\,dl =- l\,dh$, which means that both terms in eq.~(\ref{varelectroenergyflat}) are equal, hence Pelrine's doubled pressure equivalence}.

To clarify the interpretation proposed by Pelrine, let us consider the case of a \emph{compressible} elastic strip submitted to a normal stress $ {\mathcal P}/\nu$ on its surface, in the absence of any electrostatic field. We can solve Hooke's elasticity equations in the membrane using  $(\sigma_{ZZ}=- {\mathcal P}/\nu, \sigma_{XX}=T/H)$  and obtain
\begin{equation}
 \frac{EH}{1-\nu^2} e_t - \frac{1 }{1-\nu} H {\mathcal P} = T.
\end{equation}
Indeed, this is the exact same equation as (\ref{eqtanForce}), for arbitrary $\nu$. Hence, replacing all electrostatic effects by a mechanical pressure amplified by a coefficient $1/\nu$ with respect to the real electrostatic pressure exerted on a rigid plate capacitor, we recover the equation for the in-plane extension that was obtained rigorously through variational principle~\cite{Pelrine98}. This analogy %, initially proposed by Pelrine,
also provides the correct equation in cases where the strip is only partially covered by an electrode. 

However, this interpretation is not as successful when considering the normal strain $e_n$. Applying an effective pressure ${\mathcal P}/\nu$ leads to
\begin{equation}
\frac{EH}{1-\nu^2} e_n + \frac1{\nu} H {\mathcal P} = - \frac{\nu }{1-\nu}T.
\end{equation}
Comparing this to the variational result (\ref{eqenseul}), one finds that the effective pressure approach does not correctly describe the normal strain $e_n$, except for the particular limit $\nu=1/2$, i.e. if the material is incompressible. Hence, we conclude that in the planar case the effect of electrostatics can be accounted for by an enhanced normal pressure, but only in the incompressible limit. We refer to~\cite{Suo10} for a justification of this fact based on the remark that the addition of an isotropic pressure has no effect on incompressible media. 

However, even in this particular case of incompressible materials ($\nu=1/2$),  there is an additional drawback of the ``double pressure" interpretation when we turn to non planar states. Applying a pressure $2{\mathcal P}$ on both faces of a on an element of length $ds$ of elastic strip with a curvature $\kappa$,  results in a non-zero net force $2{\mathcal P} H \kappa {ds}$ acting along the normal~$\mathbf{n}$. Defining now $\mathbf{f}$ as the internal force in the strip, the equilibrium balance on an element of length $ds$ leads to $$ \frac{d\mathbf{f} }{ds} + 2{\mathcal P} H \kappa \mathbf{n} = 0 $$
 As a consequence, this would lead to a non constant internal force $\mathbf{f}$, in contrast with previous results. The torque equilibrium equation~(\ref{eqtorque}) would also be different and simply lead to $B\theta'' + \mathbf{f}.\mathbf{n} = 0 $. It does not provide the electrostatically-induced added bending stiffness of the strip $B_U$. It is therefore not exact that all electrostatic effects can be completely replaced by a double electrostatic pressure on the faces of the electrode. This approach does not extend to geometrically non-linear problems such as electrostatically induced buckling.
%Here we are rather lead to find that $\mathbf{f}+2 {\mathcal P} H\mathbf{t}$ is a constant. .
%In conclusion, the rigorous variational approach is the proper way of deriving equations and the mechanical interpretation initially proposed by Pelrine~\cite{Pelrine98} does not provide the correct distribution of forces in the system. \js{We could remove this comment and keep philosophy for the Abstract/Intro/Conclusion? Pelrine used it only for incompressible, for which it is fully correct, so no need to complain}
%\br{We have seen that some (simple) considerations : a modified electrostatic pressure (complemented with a modified bending rigidity), or a modified rest-state lead to the correct equilibrium equations for an electroactive elastic strip. The resulting equations for the beam are valid, but in both approaches, the stress distribution in the material $\sigma_{XX}$ is not correctly described. We will show in the following section the nature of the electrostatic loading, and why in particular it leads to discontinuous stresses $\sigma_{XX}$ at the electrode boundary.}

\subsection{The Maxwell stress approach} \label{MaxwellStress}

To improve upon the stress-based  interpretation, we now turn to a more systematic procedure to finding the distribution of electrostatic forces inside the elastic membrane. In continuum mechanics, electrostatic effects can be introduced via the Maxwell stress tensor~\cite{dorfmann2014}. It is defined as a tensor field whose divergence gives the electrostatic force density and reads:
\begin{equation}
\tau_{ij}^M = \varepsilon \frac{\partial \phi}{\partial x_i} \frac{\partial \phi}{\partial x_j} 
- \frac{1}{2}\varepsilon \frac{\partial \phi}{\partial x_k }  \frac{\partial \phi}{\partial x_k } \delta_{ij}. \label{eqMaxwell}
\end{equation}
Its derivation is based on the total electrostatic force on a body, via
\begin{equation}
F_i =  \int dV \, \frac{\partial \tau_{ij}^M}{\partial x_j}  = \oint dA \, \tau_{ij}^M n_j.
\end{equation}
One verifies by direct evaluation that
\begin{equation}
 \frac{\partial \tau_{ij}^M}{\partial x_j} = \varepsilon \frac{\partial^2 \phi}{\partial x_k^2 } \frac{\partial \phi}{\partial x_i} = 
 -\rho \frac{\partial \phi}{\partial x_i},
\end{equation}
where in the last step we used Gauss's law \hb{and $\rho$ is the charge density}. So indeed, the divergence of $\tau^M$ gives the expected force density $-\rho \nabla \phi$. 

From a mechanical perspective, the coupled electrostatic-elastic problem consists of solving 
\begin{equation}\label{eq:divstress}
 \frac{\partial \tau_{ij}}{\partial x_j} - \rho \frac{\partial \phi}{\partial x_i} =0, \quad 
 \mathrm{or} \quad  \frac{\partial }{\partial x_j}\left( \tau_{ij} + \tau_{ij}^M\right) =0,
\end{equation}
with appropriate boundary conditions. This provides the detailed information of the elastic stress, as well as the stress resulting from the applied electrostatic forces.

To illustrate this, we consider a flat portion of the membrane that is covered by the electrode (far away from the edge). The electrostatic field obtained in (\ref{eqpotential}) has only a component along $z$, with $\partial \phi/\partial z= U/h$. In the limit of small deformation we replace $z\simeq Z$ and $h\simeq H$, so that the Maxwell stress tensor becomes:
\begin{equation}\label{eq:maxstressplane}
\tau^M_{ZZ} = \frac{1}{2}\varepsilon \frac{U^2}{H^2}={\mathcal P}, \quad \tau^M_{XX} = - \frac{1}{2}\varepsilon \frac{U^2}{H^2}=-{\mathcal P}.
\end{equation}
Note that far away from the edges, $\tau^M$ vanishes everywhere outside the electrodes. We now  compute the total membrane tension by integrating the total stress projected in the $X$ direction, over the gap between the electrodes. This gives:
 \begin{equation} 
 T =  \int_{-H/2}^{H/2} \tau^{\mathrm{total}}_{XX} dZ=  \int_{-H/2}^{H/2} ( \tau_{XX} + \tau^M_{XX} )dZ  = H \langle \tau_{XX} \rangle - H{\mathcal P}  \label{eqMST}
 \end{equation}
 where $\langle \tau_{XX} \rangle$ is the average elastic axial stress in the membrane. Using Hooke's law and the kinematics of~(\ref{eq:membranestrain}) we do recover
 \begin{equation}  T ={\cal Y} [e_t + \frac{\nu}{1-\nu} e_n) ]-  H{\mathcal P}.
  \end{equation}
 This result is strictly identical to (\ref{eqEn}), and we see that the integral of the Maxwell stress gives the correct tension inside the membrane. We also recover, as previously deduced from energetics, that the elastic stress is discontinuous at the boundary of the electrode, since the elastic stress obeys the standard mechanical rule $\langle \tau_{XX} \rangle= T/H$ outside the electrode. We finally note that by studying Maxwell stress in curved geometries, we may also recover the electrostatic additional bending stiffness $B_U$ (see~Appendix\ref{AppendixBU}).

Given the symmetric appearance of the elastic stress $\tau$ and the Maxwell stress~$\tau^M$ in (\ref{eq:divstress}), it is tempting to give them the same interpretation -- namely, as the actual distribution of elastic and electrostatic force per unit area acting on the boundary of the body. % : When we consider an element of surface in the material with normal $\mathbf{n}$,
$\sigma \cdot \mathbf{n}$ gives the force of elastic origin acting through this surface, and in this spirit the Maxwell $\sigma^M\cdot \mathbf{n}$ would give the electrostatic force (sum of the elementary charge interaction) acting on this surface. But this is not the case.
%While this interpretation in terms of contact force is of course correct for the elastic stress, it is not straightforward for the Maxwell stress.
%because the Coulomb interaction is exerted over a distance. 
%And in fact, $\tau^M$ is defined such that it gives the correct force after integrating over the \emph{entire} area enclosing a body.%, but does this give the correct distribution of stress? In particular, does it make sense to define interpret $\tau^{\mathrm{total}} =  \tau+ \tau^M$ as a contact force per unit area?
%Still, the description in terms of Maxwell stress \br{should be taken with care}, and should not be regarded as a contact force that can be expressed as a real stress. This can be appreciated from the sketch in Fig.~\ref{paradox}, where we indicate the distribution of charges that accumulate in the electrodes at $Z=\pm H$. No free charges are present inside the membrane itself, i.e. for $-H/2 < Z < H/2$. 
For example, there are no charges in the membrane (excluding the conductors), so that no electrostatic load is directly applied through any of the surfaces -- yet, the Maxwell stresses are nonzero throughout the membrane. Maxwell stresses are therefore not the contact stresses applied by charges within the material, but provide an effective representation of long range electrostatic forces. In fact there are different possible choices of Maxwell stress~\cite{dorfmann2014}. 
\begin{figure}[b!]
\begin{center}
\includegraphics{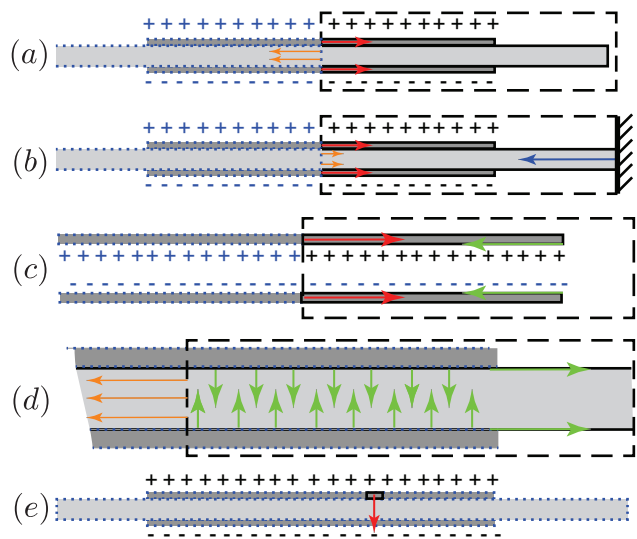} 
\end{center}
\caption{(a) Distribution of forces exerted by half of the system on the other half in a situation where no external tension is applied. (b) Same, when an external tension is applied to prevent extension \hb{(blue arrow}). (c) Electrostatic force exerted by the left part of the capacitor on the sub-system delimited by the dashed line (red arrow). The electrostatic free energy has a contribution proportional to the surface, in a close analogy with a negative surface tension. Accordingly, this energy is associated with a repulsive force in the plane of the conductors. It is balanced by an opposite force from the membrane located at the edge (green arrow). (d) Distribution of surface stress applied to the membrane (green arrows). The elastic tension results from the elastic stress applied on the left boundary (orange arrows).  (e) The normal stress exerted by the capacitor on the elastic membrane can be determined from the action reaction principle, considering a piece of conductor as the system.}
\label{paradox}
\end{figure}

\subsection{Force transmitted by surface charges} \label{mechinterp}

The actual distribution of electrostatic forces on the elastic material therefore remains to be determined. 
We start by considering the mechanical equilibrium of the volume enclosed by the dashed contour in Fig.~\ref{paradox}a, which involves both a part of the conductors and of the elastic membrane. The control surface separating the system from the external region is chosen along a normal to the plate, as is done standardly for the mechanical interpretation of surface tension. The boundary on the right is considered to be a free edge, free of any elastic or electrostatic forcing, so that the total tension $T=0$. There is a resultant electrostatic force, localised in the vicinity of the left boundary. Namely, the positive charges in the upper plate inside the control volume are subjected to a force due to surface charges in both upper and lower plates outside the control volume. However, those in the upper plate are in closer proximity, so the resultant force is dominated by the repulsive interaction -- as indicated by the red arrow. By symmetry, the same argument applies to the charges of the lower plate inside the control volume. When explicitly computing the integrals over the charge densities, one finds the force exerted on the upper (or lower) plate to be $\mathcal P H/2$ oriented to the right (cf. Appendix\ref{appendixGreen}). As $T=0$, the total repulsive electrostatic force $\mathcal P H$ must be balanced by a tensile elastic stress of equal magnitude, oriented to the left (Fig.~\ref{paradox}a, orange arrows).

Considering now the general case ($T \neq 0$ \hb{Fig.\ref{paradox}b), the total force across a section of the membrane (the tension) can be expressed as $T = - H {\mathcal P} + <\tau_{XX}>H $, where the direct electrostatic force $\frac 12 {\mathcal P} H$ complements that transmitted from one side to the other through the elastic stress:
\begin{equation}
\tau_{XX} = \frac{E}{1-\nu^2} e_t -\frac{\nu}{1-\nu} {\mathcal P}
\end{equation}
The total membrane tension $T$ consists of a bulk-elastic contribution and an \hb{electrostatic contribution localised exactly at the surface}. In Fig.~\ref{paradox}b, we exemplify the case where the length of the membrane is kept unchanged i.e. when $e_t=0$. A fraction, determined by $\nu$, of the vertical stress of electrostatic origin is redistributed horizontally, leading to an elastic contribution to the tension, due to electrostatics, but transmitted as a contact force in the bulk of the membrane (orange arrows).}

Consider now the system enclosed by the very same contour, but containing only the conductors, as shown in Fig.~\ref{paradox}c. The electrostatic  force exerted by the external region on the system is the same as before \hb{(red arrow)}. It must be balanced by an opposite force (\hb{green arrows)} exerted by the membrane on the electrode, which below we show to be located at the edge.

\subsection{Edge effects} \label{mechinterp_sing}
So far, the discussion of the electrostatic field has been simplified by avoiding the proximity of the edges of the conductor. Namely, the expressions for $\phi$ and the resulting elastic stress have been evaluated at distances from the edge large compared to the thickness $H$, avoiding fringe effects. In this thin membrane description, (\ref{eqMST}) implies a discontinuity of elastic tension $H\langle \tau_{XX}\rangle$ near the conductor's edge, since $T$ is constant while $\mathcal P$ suddenly vanishes beyond the conductor's edge. In the light of Fig.~\ref{paradox}c, it is therefore of interest to investigate the mechanics near the edge, and reveal the origin of the discontinuity.

In Fig.~\ref{paradox}d we consider the free body diagram similar to Fig.~\ref{paradox}a, but now considering the membrane only i.e. excluding the conducting plate. 
Hence, no surface charges are included into the control volume so that there are no Coulombian forces acting on the system. Instead, there are only contact forces that act at the boundary of the enclosed volume. From the action-reaction principle, one can infer that the normal stress exerted on the membrane is equal to the electrostatic stress (Fig.~\ref{paradox}e). As a consequence, turning back to Fig.~\ref{paradox}d, a compressive stress of magnitude $\mathcal P$ is exerted on the membrane in the vertical direction. In the horizontal direction, we recall that the elastic tension on the left boundary has a magnitude $H\mathcal P$, pointing to the left (orange arrow in Fig.~\ref{paradox}d). At equilibrium this horizontal force must be balanced. However, the right side is free of elastic stress and the surface charges that gave the balance in Fig.~\ref{paradox}b can no longer come to the rescue, as these are not inside the control volume. Indeed, the only possibility is that a point force emerges at the conductor edge, as previously indicated in Fig.~\ref{paradox}c. This point force is the origin of the ``discontinuity" of the elastic tension, as observed in the thin membrane description. 

\begin{figure}[t!]
\begin{center}
\includegraphics{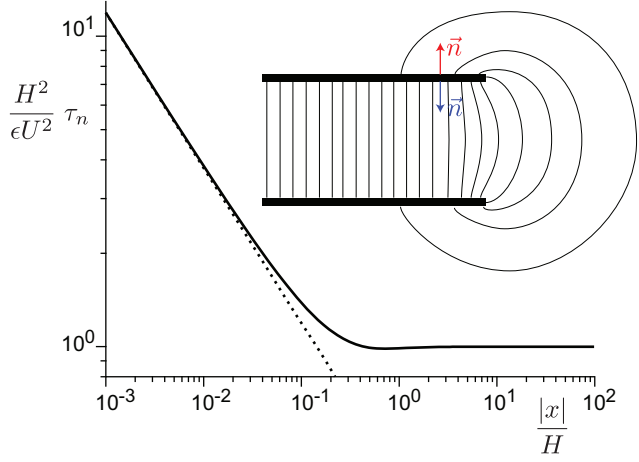} 
\end{center}
\caption{Normal stress distribution as a function of the distance to the edge (solid line). Dashed line shows the analytical asymptotics. Each piece of conductor is associated with two surface charges, one above and the other below the conductor, and accordingly, with two normal unit vectors $\vec n$.}
\label{stress}
\end{figure}

How can such a point force occur? To address this question, we again change control volume -- this time we consider only the electrostatic forces on the capacitor, excluding the elastic membrane. The electrostatic forces on the surface charges of the plate must be directly transmitted to the elastic membrane onto which the plate is glued (Fig.~\ref{paradox}e). The inset of Fig.~\ref{stress} shows the fringe field near the edge of the plate, which can be solved analytically by conformal mapping \cite{vanderlinde2006classical,valluri2000some} i.e. using a change of coordinates preserving orientation and angles locally. The idea behind this analysis is to take the solution to the infinite plane capacitor, and then to fold the boundary
by a conformal mapping to match the boundary of the finite capacitor. The derivation is given in the Appendix\ref{appendix-sec1}. Based on this analytical solution for $\phi$, we can compute the surface charge density
\begin{equation}
\sigma = -\varepsilon \nabla \phi \cdot \mathbf n,
\end{equation}
where $\mathbf n$ is the normal pointing outward the conducting plane. Even though the solution is for an infinitely thin plate, we note that one needs to distinguish between the ``inside" (or ``membrane side") of the plate ($\mathbf n$ pointing into the membrane, blue vector in Fig.~\ref{stress}) and the ``outside" ($\mathbf n$ pointing away from the membrane, red vector in Fig.~\ref{stress}). The force per unit area on the conductor will be along the normal direction and follows as
\begin{equation}
\tau_n = - \frac{1}{2} \sigma \left( \nabla \phi \cdot \mathbf n \right) = \frac{1}{2} \varepsilon |\nabla \phi|^2 = \frac{1}{2\varepsilon} \sigma^2.
\end{equation}
At the edge of the plate, where the normal $\mathbf n$ turns from upward to downward, there is indeed a singular horizontal force.  In reality, conductors are not infinitely thin so that the tangential force is not a delta function but a distributed peak. More precisely, the characteristic thickness of the conductors fixes the typical size over which the tangential force is spread over space.

Let us now discuss the results from the conformal mapping discussed in the Appendix\ref{appendix-sec1}. The exact solution for the charge density is given in a parametric form. We denote by $x$ the horizontal coordinate along the membrane and locate  the right edge at $x=0$ so that the membrane is in the half-space $x<0$. The mapping involves the following coordinate transformation
\begin{equation}
x=  \xi_\pm + \frac{H}{2\pi} \left( 1- e^{2\pi \xi_\pm/H}  \right) ,
\end{equation}
where $\xi_+$ is the new (non cartesian) coordinate for the outside part of the capacitor ($\xi_+$ varies logarithmically with $x$) and $\xi_-$ the inside part of the capacitor ($ \xi_- \simeq x$ asymptotically). The exact solution for the charge density reads
\begin{equation}
\sigma^\pm = \pm \varepsilon \frac{U}{H} \frac{1}{1-e^{2\pi \xi_\pm(x)}},
\end{equation}
where $\sigma^\pm$ is the surface charge that resides on the exterior (+) and at the interior (-) of the plate. One verifies that both inside and outside the capacitor, $\sigma$ diverges near the edge according to
\begin{equation}
\sigma \sim \varepsilon \frac{U}{H} \sqrt{\frac{H}{4 \pi |x|}}.
\end{equation}
Hence, the associated normal stress, which points in opposite directions on both sides, presents a $1/|x|$ divergence, 
\begin{equation}
\tau_n= \frac{1}{2\varepsilon} \sigma^2 \sim \varepsilon \frac{U^2}{H^2}  \frac{H}{8 \pi |x|}.
\end{equation}
One can indeed understand that such a $1/|x|$ singularity gives a finite horizontal contribution when integrating around the edge. Namely, we remind that the plate will in reality have a finite thickness that comes along with a ``rounded" edge, providing a cutoff scale $r$. In this rounded section, the normal $\mathbf n$ has a horizontal contribution. Since furthermore the length of this section is proportional to $r$, the integral over stress $\tau_n \sim \varepsilon \frac{U^2}{H r}$ will give a finite horizontal contribution $\sim \varepsilon U^2/H \sim \mathcal PH$. Hence, the point force indicated in Fig.~\ref{paradox}d is due to the integrable stress singularity at the conductor's edge.

Finally, we analyse the vertical force on the conductor. Inside the capacitor, we find that $\sigma^{-}$ decays to $\varepsilon U/h$ away from the edge while outside the capacitor, $\sigma^+$ decays to $0$. Since the normals are pointing in the opposite direction, the vertical stress $\tau_{ZZ}$ exerted on the electrode (and hence on the membrane) can be expressed as a difference of $\tau_n$ between both sides:
%%
%\begin{figure}[t!]
%\begin{center}
%\includegraphics{stress.eps} 
%\end{center}
%\caption{Normal stress distribution as a function of the distance to the edge (solid line). Dashed line shows the analytical asymptotics.}
%\label{stress}
%\end{figure}
%%
\begin{equation}
\tau_{ZZ}=\varepsilon \frac{U^2}{H^2}\left[ \frac{1}{\left(1-e^{2\pi \xi_-(x)}\right)^2}- \frac{1}{\left(1-e^{2\pi \xi_+(x)}\right)^2}\right]
\end{equation}
This result for $\tau_{ZZ}$, properly normalised, is represented in Fig.~\ref{stress} as a function of the cartesian coordinate $x$. Close to the edge, the normal stress diverges as:
\begin{equation}
\tau_{ZZ} \sim  \varepsilon \frac{U^2}{H^2} \frac{2}{3}\sqrt{\frac{H}{\pi |x|}}.
\end{equation}
This vertical force is a weaker singularity than for $\tau_n$, and does not lead to a point force at the edge upon integration. 
%%
%\begin{figure}[b!]
%\begin{center}
%\includegraphics{NegativeSurfaceTension.eps} 
%\end{center}
%\caption{Distribution of forces exerted by half of the system on the other half.}
%\label{NegativeSurfaceTension}
%\end{figure}
%%

%

\section{Concluding remarks}
We presented a theoretical study of partially active electro-membranes. We derived the equilibrium equations of an electrostatically activated beam with variational techniques. These equations were then written in an alternative way, allowing a new interpretation in terms of reference state: the electrostatic solicitation is seen as defining a new stress-free reference state. The analogy with swelling, biological growth or plastic deformation make this point of view particularly convenient when considering out-of-plane displacement and buckling problems. Indeed the difficulties arising from elastic strain discontinuities (whose origin is the singular electostatic forces at the boundary of the electrode) simply comes from a difference in rest state in this equivalent framework. We also showed that Pelrine's approach, often used, that consists of considering that the electrostatic solicitation acts as a doubled pressure along the thickness of the membrane leads to the same equation for the in plane extension. However, for polymers whose Poisson ratio differs from $1/2$, the doubled pressure does not yield the correct equation governing the normal strain. Moreover, by including geometric non-linearities in our energetic approach we shed light on  an electrostatically-induced added bending stiffness phenomenon. This feature is missing in Pelrine's approach which also generates spurious non-balanced pressure forces. In the second section we clarified the intricate balance of forces in this system. In particular we examined the question of the doubled pressure along the thickness, and gave an interpretation of the Maxwell stress tensor. Finally we considered several control contours, as is often done for surface tension problems, to understand how forces are transmitted by the charges. More precisely, by calculating the electric field at the edges of the electrode, we elucidate the major role played by the edge effects, and show that it is the origin of the elastic discontinuity of the elastic stress highlighted in the previous part.

This work is pertinent for the description of geometrically non-linear slender electro-active plates in the small strain approximation and in 2D case (elastica). It provides a better understanding of the physics of dielectric polymers that might prove useful for further developments. It remains to be generalised to more general kinematics of plates. It would be interesting to also consider the effect of large strains, which are often generated in Electro-Active devices. We however expect that the features we have presented here (e.g. singular forces at the border of the electrode leading to discontinuities in elastic strain, equivalent description in terms of a new rest state) will still be relevant in the more general case.

\section{appendix}

\subsection{A conformal mapping to solve electrostatics}
\label{appendix-sec1}
\subsubsection{Definitions}
We now zoom at a scale comparable to $H$, sending consequently the size of the capacitors to infinity. Although one can use again the Green's theorem, we will proceed through a conformal mapping. The 2D space is characterised by the complex coordinate $z=x+iy$. We define the complex potential
\begin{equation}
w(z) = \psi(x,y) + i \phi(x,y),
\end{equation}
where $\phi(z)$ the electrostatic potential. The lines of iso-$\psi$ are orthogonal to lines of iso-$\phi$, similar to the stream function and potential in ideal flow. On the domain where $w(z)$ is holomorphic, the functions $\psi$ and $\phi$ are harmonic and satisfy the Cauchy-Riemann equations. Hence, we can write
\begin{equation}
\frac{dw}{dz} = \frac{\partial \phi}{\partial y} + i \frac{\partial \phi}{\partial x}  = -  E_y - i E_x .
\end{equation}
On a conductor the density of charges reads
\begin{equation}
\sigma = - \varepsilon \nabla \phi \cdot \mathbf n,
\end{equation}
and $\nabla \phi$ is in the normal direction. Hence, the normal force per unit area can be computed as
\begin{equation}
\left( - \frac{1}{2}\sigma \nabla \phi \right)\cdot \mathbf n=  \frac{1}{2} \varepsilon |\nabla \phi|^2 =  \frac{1}{2} \varepsilon \left|\frac{dw}{dz} \right|^2.
\end{equation}

\subsubsection{Conformal mapping}
We consider the solution of the infinite plane capacitor, in the plane $\zeta=\xi+i\eta$. We allow $\eta$ to vary from $-h/2$ to $h/2$, with boundary conditions $\phi = \pm U/2$. The solution of course reads $\phi = U\eta/h$, which in terms of the complex potential gives
\begin{equation}
w(\eta) = \frac{U\zeta}{h}.
\end{equation}
One verifies $dw/d\zeta = U/h$, so as expected $E_y = -U/h$ and $E_x=0$.  Now we perform a mapping from the $\zeta$-plane to the physical plane of interest, defined by $z=x+iy$, according to \cite{Darko2012}
\begin{equation}\label{eq:map}
z =  \zeta + \frac{h}{2\pi} \left( 1 + e^{2\pi \zeta/h} \right).
\end{equation}
We verify below that this folds the conductors at $\eta=\pm h/2$ for $\xi>0$ onto the same conductors in the range $\xi <0$. Now, we use that $\zeta/h = w/U$, so that we find
\begin{equation}
\frac{z}{h} = \frac{w}{U} +  \frac{1}{2\pi}\left( 1 + e^{2\pi w/U} \right).
\end{equation}
This is a closed form solution $w(z)$ as defined by its inverse.  The properties of the field and the charge density are encoded in the derivative
\begin{equation}
\frac{dw}{dz} = \frac{dw}{d\zeta} \frac{d\zeta}{dz} = \frac{U}{h} \frac{d\zeta}{dz}.
\end{equation}
So, we can derive all results of interest from the map $z(\zeta)$ defined by (\ref{eq:map}). For simplicity we set $h=1$, so that 
\begin{equation}\label{eq:map2}
z = \zeta + \frac{1}{2\pi} \left( 1+  e^{2\pi \zeta} \right).
\end{equation}
We find that 
\begin{equation}
\frac{dz}{d\zeta} = 1 + e^{2\pi \zeta}.
\end{equation}
Singularities are encountered at when the derivative vanishes, which are found at $\zeta = \pm i/2$ (we remind that $\eta$ is restricted between -1/2 and 1/2). Indeed, these correspond to the plate edges $z=\pm i/2$. This means that the field strength (scaled on $U/h$) reads
\begin{equation}
\frac{d\zeta}{dz} =  \frac{1}{1+e^{2\pi \zeta}},
\end{equation}
and indeed diverges at the edges. To analyse the nature of the singularity, we expand the result $\zeta = i/2$, which gives
\begin{equation}
\frac{d\zeta}{dz} \sim  \frac{1}{\zeta- \frac{i}{2}}.
\end{equation}
Expanding the mapping around the point gives
\begin{equation}
z - \frac{i}{2} = - \pi(\zeta- \frac{i}{2})^2, 
\end{equation}
so that the singularity can be written as
\begin{equation}
\frac{d\zeta}{dz} \sim  \frac{1}{(z- \frac{i}{2})^{1/2}}.
\end{equation}
At the plate we can write $z=x+i/2$, so that 
\begin{equation}
\left. \frac{d\zeta}{dz}\right|_{\rm up} \sim  \frac{1}{x^{1/2}}.
\end{equation}
This implies that the field, and thus the surface charge, diverges as $1/x^{1/2}$. Next, we write the equations in terms of the real coordinates as:
\begin{eqnarray}
x &=& \xi + \frac{1}{2\pi} \left( 1+ e^{2\pi \xi} \cos 2\pi \eta \right)  \\
y &=& \eta + \frac{1}{2\pi} e^{2\pi \xi}  \sin 2\pi \eta.\label{eq:coords}
\end{eqnarray}
The upper plate is located at $\eta = 1/2$. The relation between $x$ and $\xi$ is not bijective and we well therefore note $\xi_\pm(x)$ the two branches of solution corresponding to the upper and lower side of the plate.
\begin{eqnarray}
x &=&  \xi_\pm + \frac{1}{2\pi} \left( 1- e^{2\pi \xi_\pm}  \right)  \\
y &=& \frac{1}{2}
\end{eqnarray}
For large negative values of $\xi$ one finds $ \xi_- \simeq x$ (inside capacitor), while for large positive values one finds $\xi_+ \simeq \ln(-2 \pi x) /(2\pi)$ (outside capacitor, folded from $\xi>0$).  As mentioned, the field is given by 
\begin{eqnarray}
\frac{d\zeta}{dz} =  \frac{1}{1+e^{2\pi \zeta}} = \frac{1}{1+e^{2\pi (\xi + i \eta)}},
\end{eqnarray}
which can be seen as a function of $x$ and $y$ by the inverse of (\ref{eq:coords}). At the upper plate, $\eta=1/2$, this reduces to 
\begin{eqnarray}
\frac{d\zeta}{dz} =  \frac{1}{1-e^{2\pi \xi_\pm(x)}}.
\end{eqnarray}

\subsection{Negative surface tension through the Coulomb law}\label{appendixGreen}

To determine the electrostatic force exerted on the sub-system chosen in figure~\ref{paradox}a by the rest of the electrodes, we introduce the Green's function $-\ln(r/h)/2\pi\varepsilon$ of the electrostatic interaction in the planar problem \cite{feynman1965}. 
%It gives the potential 
%
%\begin{equation}
%\phi(x,y) = - \int dx' dy' \, \sigma(x',y') \frac{1}{4\pi \epsilon} \ln\left(\frac{(x-x')^2+(y-y')^2}{h^2} \right)
%\end{equation}
%
We focus here on the infinitely extended plane capacitor, with a uniform surface charge $\pm \sigma$ at $y=\pm h/2$. Evaluated at the top electrode $n=h/2$, we find the potential
\begin{eqnarray}
\frac{U}{2} \equiv \phi(x,h/2)  &=& - \sigma \int_{-\infty}^{\infty} dx'   \frac{1}{2\pi \varepsilon} \ln\left(\frac{(x-x')^2}{h^2} \right) \nonumber \\
&& +  \sigma \int_{-\infty}^{\infty} dx'   \frac{1}{2\pi \epsilon} \ln\left(\frac{(x-x')^2+h^2}{h^2} \right) \nonumber \\
&=&  \frac{\sigma}{2\pi \varepsilon}  \int_{-\infty}^{\infty} dx'  \ln\left(\frac{x'^2+h^2}{x'^2} \right) = \frac{\sigma h }{2\epsilon}
\end{eqnarray}
We therefore recover the standard law $\sigma = \varepsilon \nabla \phi \cdot \mathbf n$, which gives $ \sigma = \varepsilon U/h$ for a plane capacitor. 
%\js{for the part below, I guess it would help to be explicit that we integrate over $d/dx$ of the potential... so we recover difference of potential for total force... no? I would guess that we can omit the explicit expressions involving $\tan^{-1}$...}

We now divide the membrane into two parts and  determine the distribution of forces exerted by one side on the other. The horizontal component of the electric field due to the other half membrane therefore reads:
\begin{equation}
\partial_x V_{\frac12}(x,y)= \frac{\sigma}{4\pi\varepsilon } \ln\left(\frac{x^2+(h-y)^2}{x^2+y^2}\right)
\end{equation}
The electric potential induced by half a capacitor as:
\begin{equation}
V_{\frac12}(x,y)=\frac{\sigma}{4\pi\varepsilon } \int_0^\infty dx' \ln\left(\frac{(x-x')^2+(h-y)^2}{(x-x')^2+y^2}\right)
\end{equation}
On the capacitor ($y=0$), we get:
\begin{equation}
V_{\frac12}(x,0)=\frac{\sigma h}{4\pi\varepsilon} \left[\frac xh \ln(1+h^2/x^2)-2 \tan^{-1}\left(\frac hx\right)\right]
\end{equation}
The horizontal repulsive force (the opposite of the tension $T_{\rm U}$) exerted by one side on the other of the form:
\begin{equation}
-T_{\rm U}=\sigma \left[V_{\frac12}(0,0)-V_{\frac12}(\infty,0)\right]= \frac{\sigma^2 H}{4\varepsilon }=\frac{\varepsilon}{4 } \frac{U^2}{H}=\frac{{\mathcal P}H}{2}
\end{equation}

\subsection{Electrostatic bending stiffness computed from Maxwell stress tensor} \label{appendixBMaxwell} \label{AppendixBU}
We may also compute the effective bending stiffness from electrostatic origin using Maxwell stress tensor defined in (\ref{eqMaxwell}). 
From   the potential computed in a curved capacitor (\ref{eqpotential}), we deduce the Maxwell stress tensor component
\begin{equation}
\tau^M_{XX}=- \frac{\varepsilon}{2} \frac{ U^2}{h^2} \left[1 + Z\kappa \right]^2 = - \frac{\varepsilon}{2} \frac{ U^2}{h^2} \left[1 +2 Z\kappa \right]
\end{equation}
at linear order in $\kappa$.
This lead to a total torque (per unit width)
\begin{equation}
 M=\int_{Z=-h/2}^{Z=h/2} \left( \sigma_{XX}  + \tau^M_{XX}  \right) Z  dZ   = - B\kappa  - {\varepsilon} \frac{ U^2}{h^2}  \left[ \frac{h^3}{12} \kappa \right] = - (B + B_U)  \kappa, 
\end{equation}
where we have computed the elastic stress $\sigma_{XX}$ from Hooke's law and the strains defined in (\ref{eq:membranestrain}).
We have therefore recovered the electrostatically-induced additional bending stiffness $B_U$ with the same expression as in (\ref{eqBu}).

\bibliographystyle{elsarticle-num}

\bibliography{DielectricEelastomer,sample}

\end{document}